\documentclass[aps,showpacs,twocolumn]{revtex4}
\usepackage{amssymb}
\usepackage{amsmath}
\usepackage{graphicx}
\usepackage{epsfig}

\begin{document}

\title{Long-range adiabatic quantum state transfer through tight-binding
chain as quantum data bus}
\author{Bing~Chen$^{1}$, Wei~Fan$^{1}$, Yan~Xu$^{1,3}$, Zhao-yang~Chen$^{2}$%
, Xun-li~Feng$^{3}$ and C. H. Oh$^{3}$}
\affiliation{$^{1}$School of Science, Shandong University of Science and Technology,
Qingdao 266510, China}
\affiliation{$^{2}$ Department of Mechanical Engineering, University of California,
Berkeley, CA 94720, USA}
\affiliation{$^{3}$ Center for Quantum Technologies and Physics Department, Faculty of
Science, National University of Singapore, 2 Science Drive 3, Singapore
117542}
\date{\today}

\begin{abstract}
We introduce a scheme based on adiabatic passage which allows for long-range
quantum communication through tight-binding chain with \emph{always-on}
interaction. By adiabatically varying the external gate voltage applied on
the system, the electron can be transported from the sender's dot to the aim
one. We numerically solve the schr\"{o}dinger equation for a system with
given number of quantum dots. It is shown that this scheme is a simple and
efficient protocol to coherently manipulate the population transfer under
suitable gate pulses. The dependence of the energy gap and the transfer time
on system parameters is analyzed and shown numerically. Our method provides
a guidance for future realization of adiabatic quantum state transfer in
experiments.
\end{abstract}

\maketitle

\section{Introduction}

Quantum state transfer (QST), as the name suggests, refers to the transfer
of an arbitrary quantum state from one qubit to another, which is a central
task in quantum information science. For the solid-state based quantum
computing at the large scale, it is very crucial to have a solid system
serving as such quantum data bus, which can provide us with a quantum
channel for quantum communication. During the last years many efforts have
been made in different fields to design a feasible proposal for perfect QST.
One kind of proper QST proposals is based on solid-state system with \emph{%
always-on} interaction~\cite{Bose1,Song,Christandle1} . The communication is
achieved by simply placing a quantum state at one end of the chain and
waiting for an optimized time to let this state propagate to the other end
with a high fidelity. The other kinds of proposals have paid much attention
to adiabatic passage for coherent QST in time-evolving quantum systems,
which is a powerful tool for manipulating a quantum system from an initial
state to a target state. This way of population transfer has the important
property of being robust against small variations of the Hamiltonian and the
transport time, which is crucial in experiment since the system parameters
are often hard to control. The typical scheme for coherently spatial
population transfer has been independently proposed for neutral atoms in
optical traps~\cite{Eckert} and for electrons in quantum dot (QD) systems~%
\cite{CTAP} via a dark state of the system, which is termed coherent
tunneling via adiabatic passage (CTAP) following Ref.~\cite{CTAP}. In such a
scheme, the tunneling interaction between adjacent quantum units is
dynamically tuned by changing either the distance or the height of the
neighboring potential wells following a counterintuitive scheme which is a
solid-state analog of the well-known stimulated Raman adiabatic passage
(STIRAP) protocol~\cite{STIRAP} of quantum optics. Since then, the CTAP
technique has been proposed in a variety of physical systems for
transporting single atoms~\cite{atom1,atom2}, spin states~\cite{spin},
electrons~\cite{electron1,electron2} and Bose-Einstein condensates~\cite%
{BEC1,BEC2,BEC3}. It has also been considered as a crucial element in the
scale up to large quantum processors~\cite{LR1,LR2}.

Recently, Ref.~\cite{chen1} presented a scheme to adiabatically transfer an
electron from the left end to the right end of a three dot chain using the
ground state of the system. This technique is a copy of the frequency
chirping method~\cite{CF1,CF2}, which is used in quantum optics to transfer
the population of a three-level atom of the Lambda configuration. The scheme~%
\cite{chen1} is presented as an alternative to a well known transfer scheme
(CTAP)~\cite{CTAP}. However, different from CTAP process, the protocol in
Ref.~\cite{chen1} considers a three QD array with \emph{always-on}
interaction which can be manipulated by the external gate voltage applied on
the two external dots (sender and receiver). Through maintaining the system
in the ground state, it shows that it is a high-fidelity process for a
proper choice of system parameters and also robust against experimental
parameter variations. The obvious extension of this work is to consider the
passage through more than one intervening dot~\cite{chen2}. In this paper we
will consider a quasi-one-dimensional chain of QDs, which is schematically
illustrated in Fig.~\ref{fig1}. The central tight-binding chain serves as
the quantum channel and two external QDs are attached to the media chain.
The sender (Alice) and the receiver (Bob) control one external QD each and
Alice transmits information via a qubit using the chain to Bob by
adiabatically changing the gate voltages. Different from previously
discussed schemes, we will consider a fixed $N$-site coupled QDs media chain
and QST can be realized in required transfer distance by modulating the
positions where QD A and B are connected to the chain. In particular, the
nearest-neighbor hopping amplitudes are set to be uniform. We first
theoretically elaborate the adiabatic QST in this scheme. Taking a 50-dot
structure as an example, we show that the electron can be robustly
transported from Alice to Bob through the media chain, by slowly varying the
gate voltages.

The paper is organized as follows. In Sec. II the model is setup and we
describe the adiabatic transfer of an electron between QDs. In Sec. III we
show numerical results that substantiate the analytical results. The last
section is the summary and discussion of the paper.

\section{Model Setup}

\begin{figure}[tbp]
\includegraphics[ bb=103 157 450 501, width=7 cm, clip]{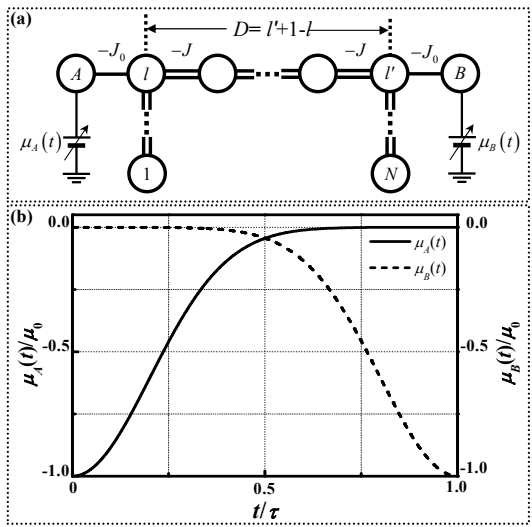}
\caption{(a) Schematic illustrations of adiabatic QST in multi-dot array.
The system is controlled by external gates voltage $\protect\mu _{i}(t)$ $%
(i=A,$ $B)$. By adiabatically varying the gates voltage, one can achieve
long-range QST from the QD $A$ to QD $B$ of the chain. (b) Gate voltages as
a function of time (in units of $\protect\tau $). $\protect\mu _{A}(t)$ is
the solid line and $\protect\mu _{B}(t)$ is the dash line.}
\label{fig1}
\end{figure}
Consider a quasi-one-dimensional chain of QDs, realized by the empty or
singly occupied states of a positional eigenstate, see Fig.~\ref{fig1}. The
whole quantum system consists of two sites ($A$ and $B$) and a simple
tight-binding $N$-site chain. The sender (Alice) and the receiver (Bob) can
only control the external gates voltage $\mu _{\alpha }(t)$ $(\alpha =A,$ $%
B) $. The total Hamiltonian $\mathcal{H}(t)=\mathcal{H}_{M}+\mathcal{H}_{I}+%
\mathcal{H}_{C},$contains three parts, the medium Hamiltonian
\begin{equation}
\mathcal{H}_{M}=-J\sum_{j=1}^{N-1}\left\vert j\right\rangle \left\langle
j+1\right\vert +\text{h.c.}
\end{equation}%
describing the tight-binding chain with uniform nearest neighbor hopping
integral $-J$ $(J>0),$ the coupling Hamiltonian
\begin{equation}
\mathcal{H}_{I}=-J_{0}\left( \left\vert A\right\rangle \left\langle
l\right\vert +\left\vert B\right\rangle \left\langle l^{\prime }\right\vert
\right) +\text{h.c.},
\end{equation}%
describing the connections between QDs $A$, $B$, and the chain with hopping
integral $-J_{0}$ $(J_{0}>0)$, and the operating Hamiltonian
\begin{equation}
\mathcal{H}_{C}=\mu _{A}(t)\left\vert A\right\rangle \left\langle
A\right\vert +\mu _{B}(t)\left\vert B\right\rangle \left\langle B\right\vert
\end{equation}%
describing the adiabatic manipulation of the Hamiltonian parameters. In $%
\mathcal{H}(t)$, $\left\vert j\right\rangle $ represents the Wannier state
localized in the $j$-th quantum site for $j=A,$ $1,$ $2,...,$ $N,$ $B$. In $%
\mathcal{H}_{I}$, $l$ and $l^{\prime }$ denote the sites of medium
connecting to the QDs $A$ and $B$. The distance between $A$ and $B$ is $%
D=l^{\prime }+1-l$. In this proposal, we just consider one connection way: $%
l^{\prime }=N+1-l$, that is the quantum state is transferred between site $A$
and its \emph{mirror-conjugate} site $B$. In term $\mathcal{H}_{C}$, $\mu
_{A}(t)$ and $\mu _{B}(t)$ are site energies (externally controlled), which
are modulated by a Gaussian pulses (shown in Fig.~\ref{fig1}(b))

\begin{eqnarray}
\mu _{A}(t) &=&-\mu _{0}\exp \left[ -\frac{1}{2}\alpha ^{2}t^{2}\right] ,
\notag \\
\mu _{B}(t) &=&-\mu _{0}\exp \left[ -\frac{1}{2}\alpha ^{2}\left( t-\tau
\right) ^{2}\right] ,  \label{miu}
\end{eqnarray}%
where $\mu _{0}$ is the peak voltage of the pulse; $\tau $ and $\alpha $ the
total adiabatic evolution time and standard deviation of the control pulse.
To realize high fidelity transfer in this scheme, the peak voltage $\mu _{0}$
must be much larger than hopping integral, i.e. $\mu _{0}\gg J,J_{0}$. The
reason is that small peak values improve adiabaticity, but lead to a low
fidelity because the final instantaneous eigenstate is not the desired one.
According to Ref.~\cite{chen2}, the transfer is optimized when we choose $%
\alpha =8/\tau $. Throughout this paper, all energies ($J_{0}$ and $\mu _{0}$%
) are scaled in units of $J$, and evolution time $\tau $ is in units of $1/J$%
.

In this proposal, we focus our study on the ground state $\left\vert \psi
_{g}(t)\right\rangle $ of Hamiltonian $\mathcal{H}(t)$ to induce population
transfer from state $\left\vert A\right\rangle $ to $\left\vert
B\right\rangle $. For single electron transfer,\ a state in the single
particle Hilbert space is assumed as $\left\vert \psi _{k}\right\rangle
=f_{A}^{k}\left\vert A\right\rangle +\sum_{j=1}^{N}f_{j}^{k}\left\vert
j\right\rangle +f_{B}^{k}\left\vert B\right\rangle $, where $k$ denotes the
momentum. Duo to the translational symmetry of the present system, the
instantaneous Hamiltonian's eigen equation for $f_{j}^{k},$ $j\in \left[ 1,N%
\right] $ is easily shown to be
\begin{widetext}
\begin{eqnarray}
-J\left[ f_{j-1}^{k}+f_{j+1}^{k}\right] &=&\left[ \varepsilon
_{k}-V_{A}\left( \varepsilon _{k}\right) \delta _{j,l}-V_{B}\left(
\varepsilon _{k}\right) \delta _{j,N+1-l}\right] f_{j}^{k},  \notag \\
J_{0}f_{A}^{k} &=&V_{A}\left( \varepsilon _{k}\right) f_{l}^{k},  \label{EQ}
\\
J_{0}f_{B}^{k} &=&V_{B}\left( \varepsilon _{k}\right) f_{N+1-l}^{k},  \notag
\end{eqnarray}%
\end{widetext}here $\varepsilon _{k}$ is the eigenenergy and the term
\begin{equation}
V_{i}\left( \varepsilon _{k}\right) =\frac{J_{0}^{2}}{\varepsilon _{k}+\mu
_{i}},i\in \left[ A,B\right]
\end{equation}%
on the right hand side is contributed by the interactions between the sites $%
A$, $B$, and medium chain which is dependent on the eigenenergy $\varepsilon
_{k}$. The $\delta $-type potential forms a confining barrier to the
transportation of single electron in the chain and forms a bounded state of
single electron, similar to those proposed in Ref.~\cite{Sun}. In this
work, we focus our attention on the bound state, which is the ground state
of the total system to realize the long-range QST.

Starting from $t=0$, we have $\mu _{A}(0)=-\mu _{0}$ and $\mu _{B}(0)\approx
0$. The solution to Eq.~(\ref{EQ}) is
\begin{widetext}
\begin{equation}
\left\vert \psi _{g}(0)\right\rangle =\mathcal{N}^{-1/2}\left[
\sum_{j=1}^{N}e^{-\kappa \left\vert j-l\right\vert }\left\vert
j\right\rangle +\frac{\lambda _{+}}{J_{0}}\left\vert A\right\rangle +\frac{%
\lambda _{-}}{J_{0}}e^{-\kappa (N+1-2l)}\left\vert B\right\rangle \right] ,
\label{psi_g}
\end{equation}
\end{widetext}where $\lambda _{\pm }=\left( \sqrt{\mu _{0}^{2}+4J_{0}^{2}}%
\pm \mu _{0}\right) /2$ and $\kappa =\ln \left( \lambda _{+}/J\right) $; $%
\mathcal{N}=\sum_{j=1}^{N}e^{-2\kappa \left\vert j-l\right\vert }+\left( \mu
_{0}^{2}+2J_{0}^{2}\right) /J_{0}^{2}$ is the normalization factor. By
choosing a sufficiently large value of $\mu _{0}$, the ground state $%
\left\vert \psi _{g}(0)\right\rangle $\ can be reduced to $\left\vert \psi
_{g}(0)\right\rangle \approx \left\vert A\right\rangle $.

With the same results, in the time limit $t=\tau $, the parameter $\mu
_{A}(t)$ goes to zero and $\mu _{B}(t)$ goes to $-\mu _{0}$. Duo to the
reflection symmetry (relabeling sites from right to left) of the system, for
$j=A,1,2,...,N,B,$
\begin{equation}
\langle j\left\vert \psi _{g}(\tau )\right\rangle =\langle \bar{j}\left\vert
\psi _{g}(0)\right\rangle .
\end{equation}%
We have used $\bar{j}=N+1-j$ to indicate the \emph{mirror-conjugate} site of
$j$. This leads to the final ground state $\left\vert \psi _{g}(\tau
)\right\rangle \approx \left\vert B\right\rangle $. To illustrate with an
example, the probability of $\left\vert B\right\rangle $ in $\left\vert \psi
_{g}(\tau )\right\rangle $ can achieve 99.7\% when the parameters are set to
be $J_{0}/J=1$ and $\mu _{0}/J=20$. Preparing the system in state $%
\left\vert \Psi \left( t=0\right) \right\rangle =\left\vert A\right\rangle
\, $and adiabatially changing $\mu _{A}(t)$ and $\mu _{B}(t)$, one can see
that the system will end up in $\left\vert B\right\rangle $,

\begin{equation}
\left\vert \Psi \left( t=0\right) \right\rangle =\left\vert A\right\rangle
\rightarrow \left\vert \Psi \left( t=\tau \right) \right\rangle =\left\vert
B\right\rangle ,
\end{equation}%
hence we can see that a high fidelity transfer may still be possible, even
with imperfect controls.

\begin{figure}[tbp]
\includegraphics[ bb=62 71 360 294, width=7 cm, clip]{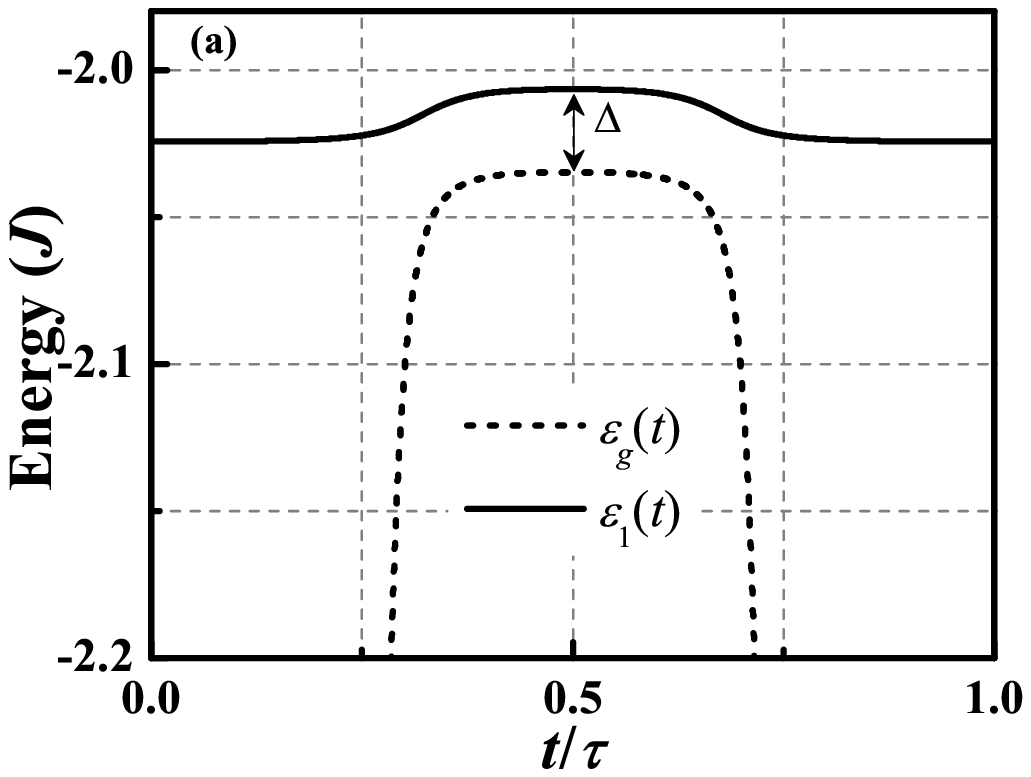} %
\includegraphics[ bb=77 226 389 716, width=7 cm, clip]{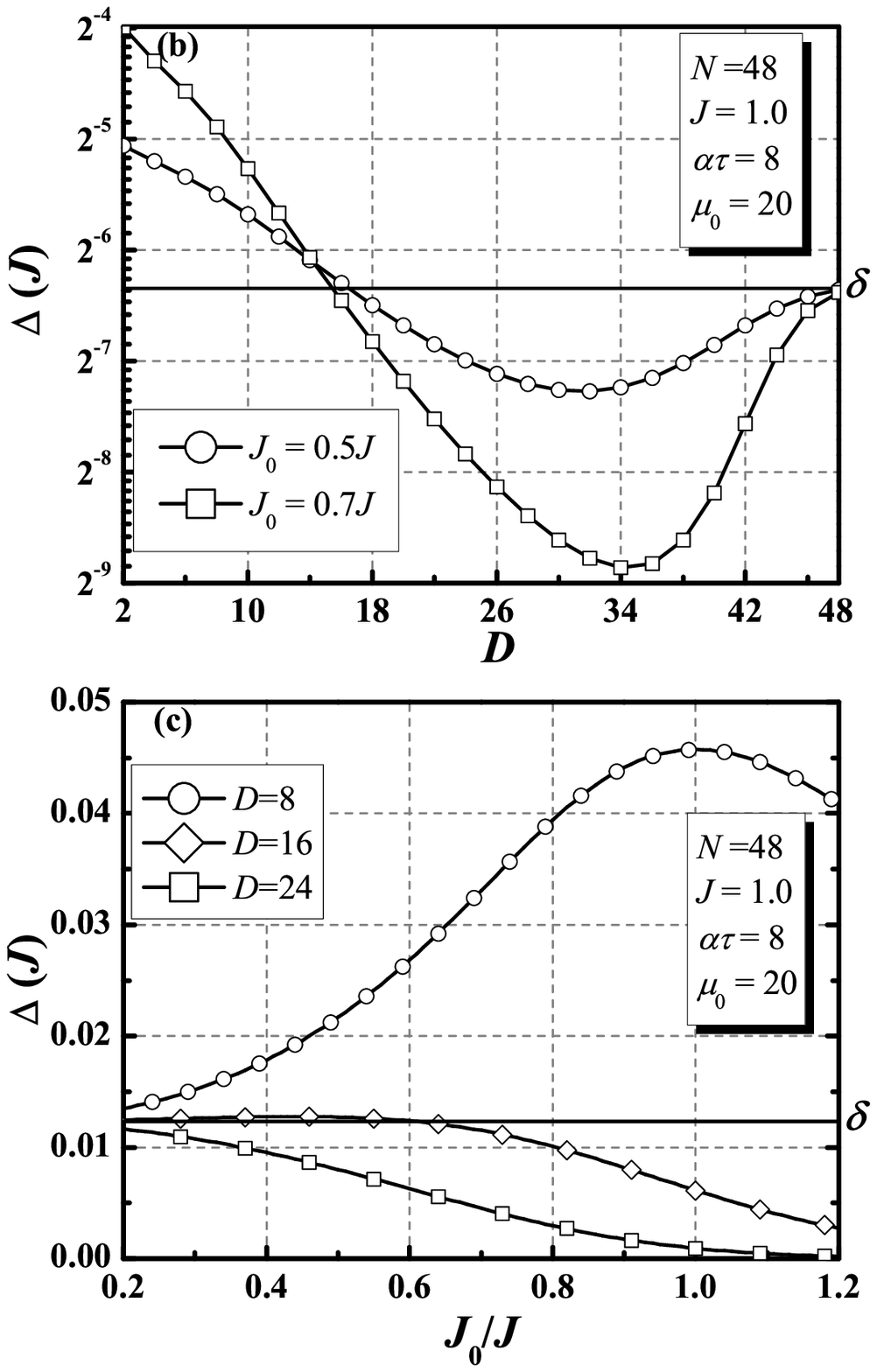}
\caption{(a) The instantaneous eigenenergy (in units of $J$) of the lowest two states $\protect%
\psi_{g}$ and $\protect\psi_{1}$ through the gate pulse shown in Fig.~1(b).
The gap is minimum at $t=\protect\tau/2$, $\Delta =\protect\varepsilon %
_{1}\left( \protect\tau /2\right) -\protect\varepsilon _{g}\left( \protect%
\tau /2\right) $. (b) Plot of gap $\Delta$ (in units of $J$) as a function of transfer
distance $D$ for $J_{0}=$0.5$J$, and 0.7$J$. The energy gap $\Delta$ is
exponentially decreasing as transfer distance $D$ increase for a given $%
J_{0} $. (c) The gap $\Delta$ (in units of $J$) as a function of coupling constant $J_{0}$ for
the transfer distance $D=$8, 16, and 24. The horizontal line $\protect\delta$
is the minimum gap of the medium chain $\mathcal{H}_{M}$. }
\label{fig2}
\end{figure}

In the absence of hopping term between the two external QDs $A$, $B$ and the
medium chain ($J_{0}=0$), the Hamiltonian $\mathcal{H}(t)$ can be
diagonalized as
\begin{equation*}
\mathcal{H}(t)=\sum_{k}-2J\cos k\left\vert k\right\rangle \left\langle
k\right\vert +\mu _{A}(t)\left\vert A\right\rangle \left\langle A\right\vert
+\mu _{B}(t)\left\vert B\right\rangle \left\langle B\right\vert
\end{equation*}%
with $\left\vert k\right\rangle =\sqrt{2/\left( N+1\right) }\sin
(kj)\left\vert j\right\rangle $, where $k=n\pi /(N+1),n=1,2,...,N$. One can
see that at $t=\tau /2$, the eigenstates $\left\vert A\right\rangle $ and $%
\left\vert B\right\rangle $ are degenerate which leads to a breakdown of
adiabaticity. The presence of the hopping term $J_{0}$ will open up energy
gap at the crossing. To evaluate instantaneous eigenvalues of the
Hamiltonian is generally only possible numerically. In Fig.~\ref{fig2}(a) we present
the results showing the eigenenergy gap between the instantaneous
first-excited state and ground state undergoing evolution due to modulation
of the gate voltages according to pulse Eq.~(\ref{miu}). The eigenvalues
shown in this figure exhibit pronounced avoided crossing and approach
nonzero minimum $\Delta =\varepsilon _{1}\left( \tau /2\right) -\varepsilon
_{g}\left( \tau /2\right) $. This minimum energy gap plays a significant
role in the transfer, because the total evolution time should be larger
enough compared to $\Delta $. In this scheme, the energy gap $\Delta $
depends both on the transfer distance $D$ and the coupling constance $J_{0}$%
. To study the relationship between the total evolution time and system
parameters is one of the important contributions of this paper.

As an example, we show in Fig.~\ref{fig2} the effect two factors has on the energy
gap $\Delta $ for a system with $N=48$ QDs and coupling strength $J=1.0$.
The pulse's parameters we choose are $\alpha =8/\tau $, and $\mu _{0}=20$.
In Fig.~\ref{fig2}(b), we plot the energy gap $\Delta $ as a function of transfer
distance $D$ for $J_{0}=0.5J$, and $J_{0}=0.7J$. The horizontal line $\delta
\approx 3J\pi ^{2}/N^{2}$ indicates the minimum gap of medium chain $%
\mathcal{H}_{M}$. One can see that the logarithmic scales chosen suggest
that as transfer distance $D$ increasing there is an exponential
disappearance of the gap $\Delta $. The smaller hopping constant $J_{0}$,
the slower the decay of $\Delta $. The other thing is that $\Delta $ is also
determined by the coupling strength $J_{0}$. Fig. 2(c) shows the numerically
computed behavior of $\Delta $ as a function of $J_{0}$ for $D=8$, $16$, and
$24$. As $J_{0}$ increases, the gap $\Delta $ increase for short-distance
transfer ($D<N/3$) and decreases for long-distance transfer ($D>N/3$). The
results shown in Fig.~\ref{fig2}(c) also suggest that decreasing coupling strength $%
J_{0}$ can obtain relatively large $\Delta $ for long-range transfer. But it
does not mean the weaker the coupling $J_{0}$,\ the better the result of QST
will be. The reason is that the energy gap is not the sufficient and
necessary condition for adiabatic process. In this proposal, the negative
effects of distance on the gap can be partially compensated by $J_{0}$.

\section{Numerical Simulations}

In this section let us firstly review the transfer process of this protocol.
At $t=0$ we initialize the device so that the electron occupies site-$1$,
i.e., the total initial state is $\left\vert A\right\rangle $. Provided we
transform the gate pulses adiabatically, then the adiabatic theorem states
that the system will stay in the same eigenstate. Therefore, the quantum
state starting in $\left\vert A\right\rangle $ will end up in $\left\vert
B\right\rangle $.

\begin{figure}[tbp]
\includegraphics[ bb=83 136 348 712, width=7 cm, clip]{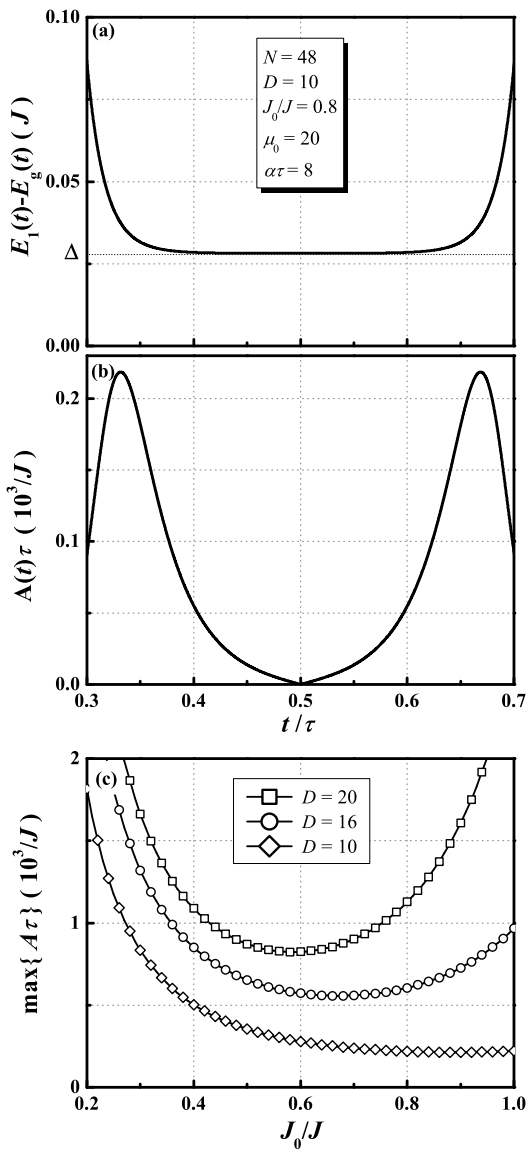}
\caption{(a) The energy difference $E_{1}(t)-E_{g}(t)$ (in units of $J$) and (b) adiabaticity $%
\mathcal{A}\left( t\right)\protect\tau$ (in units of $10^{3}/J$) as a function of time in the time
interval $t\in [0.3\protect\tau, 0.7\protect\tau]$. Note that the energy gap
approximately equals $\Delta$ in a wide time range and the maxima of
adiabaticity is not at the middle of the pulse sequence. (c) Maximum
adiabaticity $\max \{\mathcal{A}\left( t\right) \protect\tau \}$ through the
protocol as a function of $J_{0}$ for three different transfer distances. $%
\max \{\mathcal{A}\left( t\right) \protect\tau \}$ varies with $J_{0}$, and
reaches minimum value when the $J_{0}$ takes specific value.}
\label{fig3}
\end{figure}

The analysis above is based on the assumption that the adiabaticity is
satisfied. The adiabaticity parameter defined for this scheme is
\begin{equation}
\mathcal{A}\left( t\right) =\frac{\left\vert \left\langle \psi
_{g}(t)\right\vert \partial \mathcal{H}/\partial t\left\vert \psi
_{1}(t\right\rangle \right\vert }{\left\vert \varepsilon _{g}(t)-\varepsilon
_{1}(t)\right\vert ^{2}},
\end{equation}%
where $\left\vert \psi _{g}(t\right\rangle $ ($\left\vert \psi
_{1}(t\right\rangle $) is the instantaneous ground state (first-excited
state) of the Hamiltionian $\mathcal{H}(t)$ and $\varepsilon _{g}(t)$ ($%
\varepsilon _{1}(t)$) is the corresponding instantaneous eigenvalue of the
state $\left\vert \psi _{g}(t\right\rangle $ ($\left\vert \psi
_{1}(t\right\rangle $). For adiabatic evolution of the system we require $%
\mathcal{A}\left( t\right) \ll 1$ for all time, which greatly suppresses the
quantum transition from the ground state $\left\vert \psi
_{g}(t)\right\rangle $ to the first-excited state $\left\vert \psi
_{1}(t)\right\rangle $. Fig.~\ref{fig3}(a) and (b) show the numerical result of energy
difference $\varepsilon _{1}(t)-\varepsilon _{g}(t)$ and $\mathcal{A}\left(
t\right) \tau $ as a function of pulse time. Note that the appearance time
of maxima of $\mathcal{A}\left( t\right) $ is not at the middle of the
pulse sequence, but the energy difference $\varepsilon _{1}(t)-\varepsilon
_{g}(t)$ at this time is slightly larger than the minimum gap $\Delta $. So
we can use minimum gap $\Delta $ to estimate the minimum pulse time
required for high-fidelity transfer. Fig.~\ref{fig3}(c) shows the maximum
adiabaticity $\max \{\mathcal{A}\left( t\right) \tau \}$ through the
protocol as a function of $J_{0}$ for $D=10$, $16$, and $20$. One sees that
there is an optimal value of $J_{0}$\ which ensures the shortest time for
realizing perfect QST and the optimal value decreases as transfer distance $%
D $ increases. To sum up, the adiabatic regime necessary to obtain transport
with high fidelity can be concluded to the condition $\tau \gg 1/\Delta
^{\ast }\left( D\right) $, where $\Delta ^{\ast }\left( D\right) $ is the
minimum gap of the system when the coupling strength $J_{0}$ takes the
optimal value corresponding to transfer distance $D$.

The consequent time evolution of the state is given by the Schr\"{o}dinger
equation (assuming $\hbar =1$)
\begin{equation}
i\frac{d}{dt}\left\vert \Psi \left( t\right) \right\rangle =\mathcal{H}%
(t)\left\vert \Psi \left( t\right) \right\rangle .  \label{SEQ}
\end{equation}%
The time evolution creates a coherent superposition:%
\begin{equation}
\left\vert \Psi \left( t\right) \right\rangle =c_{A}(t)\left\vert
A\right\rangle +\sum_{j=1}^{N}c_{j}(t)\left\vert j\right\rangle
+c_{B}(t)\left\vert B\right\rangle ,
\end{equation}%
where $c_{j}(t)$ denotes the time-dependent probability amplitude for the
electron to be in $j$-th QD. We define the probability of finding the
electron on the medium chain as $\left\vert c_{M}(t)\right\vert
^{2}=\sum_{j=1}^{N}\left\vert c_{j}(t)\right\vert ^{2}$ that obeys the
normalization condition$\,\left\vert c_{A}(t)\right\vert ^{2}+\left\vert
c_{M}(t)\right\vert ^{2}+\left\vert c_{B}(t)\right\vert ^{2}=1$. At time $%
\tau $ the fidelity of initial state transferring to the dot-$B$ is defined
as $F=\left\vert \langle B\left\vert \Psi (\tau )\right\rangle \right\vert
^{2}=\left\vert c_{B}(\tau )\right\vert ^{2}$.

\begin{figure*}[t]
\includegraphics[ bb=50 33 378 795, width=7 cm, clip]{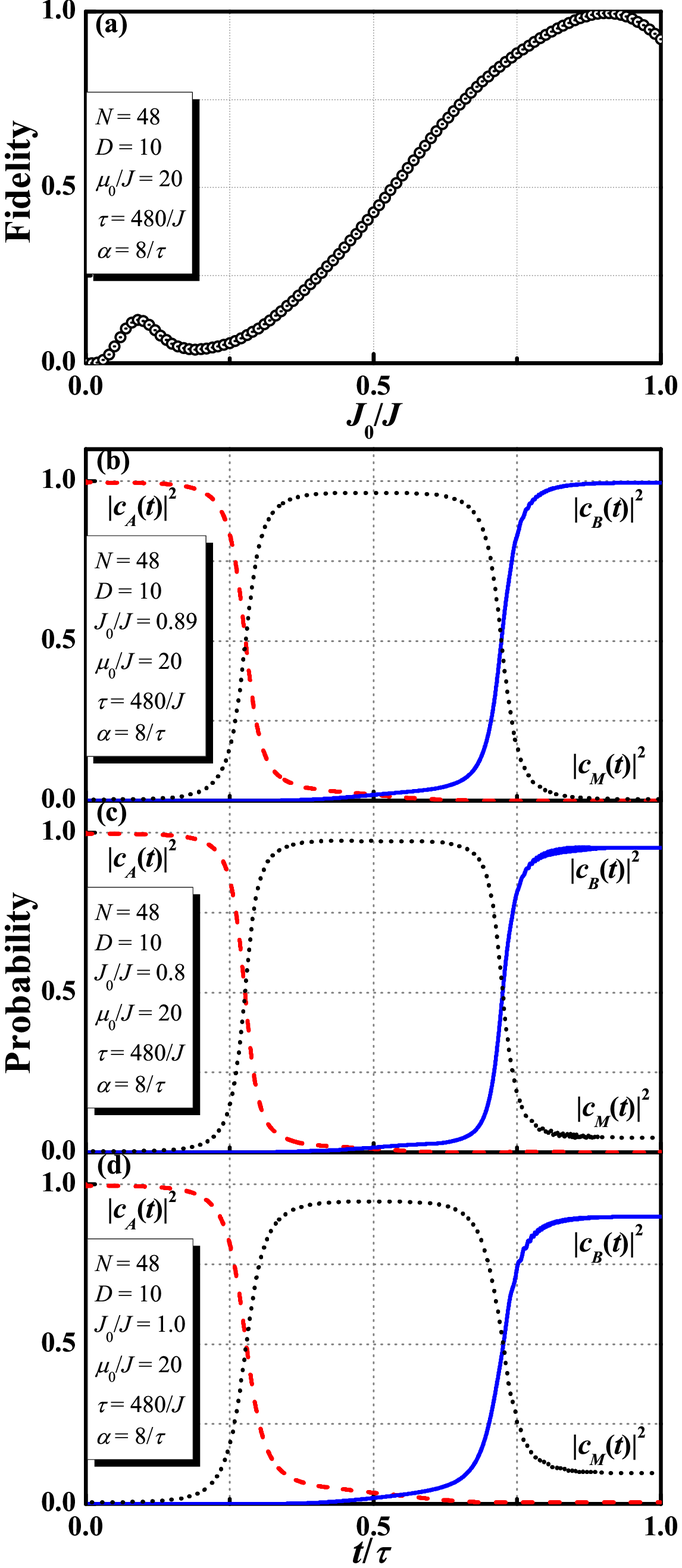} %
\includegraphics[ bb=50 33 378 795, width=7 cm, clip]{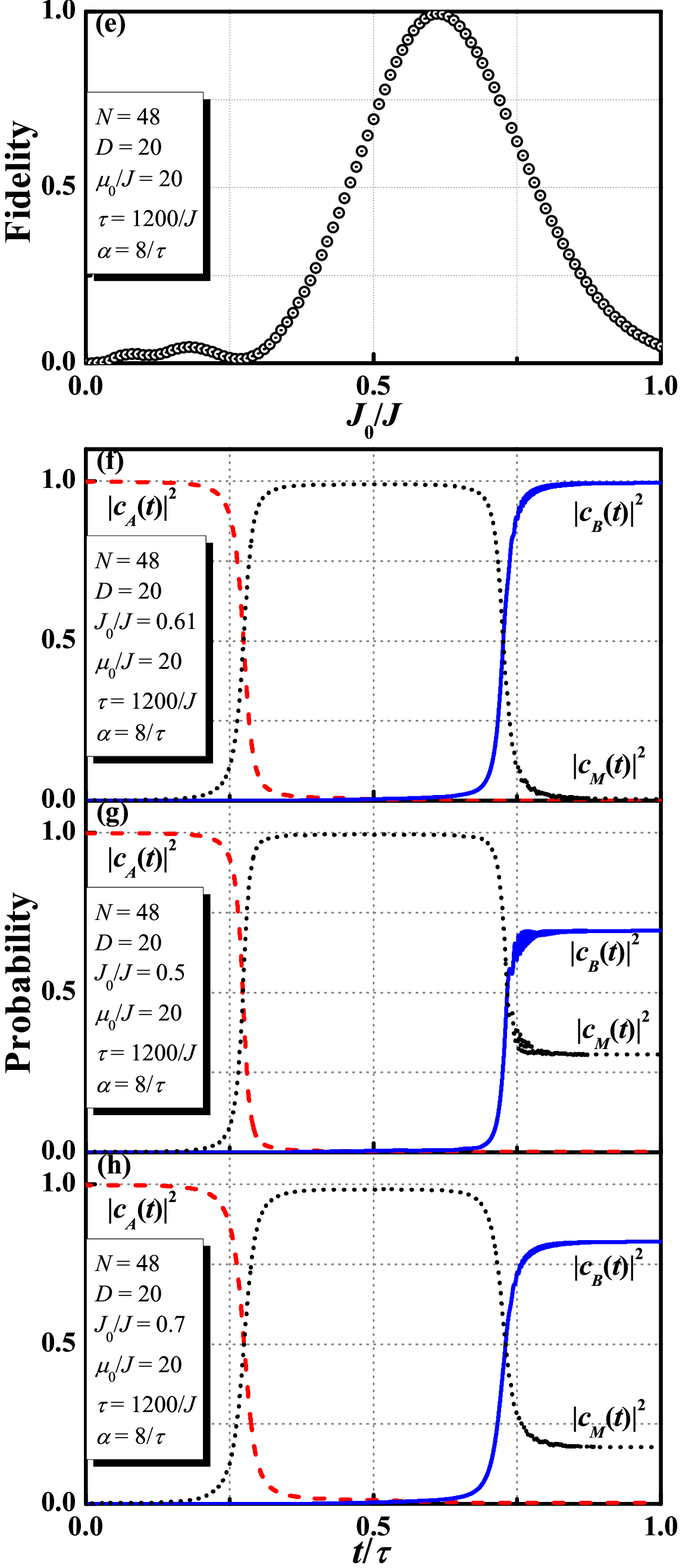}
\caption{(Color online) QST from QD $A$ to $B$ attached to a tight-binding
chain of length with $N=48$ for two different transfer distance: $D=11$
(a)-(d) and $D=21$ (e)-(h). (a) The transfer fidelity as a function of
coupling strength $J_{0}$ for transfer distance $D=11$. The calculation was
done for various values of $J_{0}$ at fixed total evolution time $\protect%
\tau =480/J$. Note that there is an optimal value of $J_{0}$ to achieve high
fidelity QST. Population transfer as a function of time obtained for three
coupling strength $J_{0}$:(b) $J_{0}=0.8J$; (c) $J_{0}=0.91J$; (d) $%
J_{0}=1.0J$. Initially the population is on QD $A$ (dashed red line) and finally
mainly on QD $B$ (solid blue line). The population on the media chain is shown as
a dotted black line. (e) The same as in (a), but for transfer distance $D=21$ and $%
\protect\tau =1200/J$. (f)-(h) The population behavior under the influence
of $J_{0}$. The parameters are like that in (b)-(d).}
\label{fig4}
\end{figure*}

In order to proceed, we used standard numerical methods to integrate the Schr%
\"{o}dinger equation for probability amplitudes. Because the scheme relies
on maintaining adiabatic conditions, we examine the effect of system
parameters on the target state population. In Fig.~\ref{fig4}, we consider the system
with $N=48$ and show QST from QD $A$ to QD $B$ for two different transfer
distance: $D=11$ and $D=21$. Firstly, we examine the effect of coupling
strength $J_{0}$ on transfer fidelity. It is seen that there is an optimal
value of $J_{0}$ to achieve high-fidelity transfer.

To illustrate the process of QST for $D=11$, we exhibit in Fig.~\ref{fig4}(b)-(d) the
exact evolution of the probabilities of finding electron in QD $A$ (red dashed
line), $B$ (blue solid line), and media chain (black dotted line) as a function of time
for three different values of coupling strength $J_{0}$ but the same
remaining parameters ($\mu _{0}=20J$, $\tau =480/J$ and $\alpha =8/\tau $).
We get good results for the transfer if we choose $J_{0}=0.89J$ as shown in
Fig.~\ref{fig4}(b). The populations on the QD $A$ and QD $B$ are exchanged in the
expected adiabatic manner. If we choose parameters deviated from the optimum
value, $J_{0}=0.8J$ and $1.0J$, we find the results in Fig.~\ref{fig4}(c) and Fig.~
\ref{fig4}(d). We can see that a slight deviations from the values will breaks
adiabaticity and lead to major deteriorations of the quality of transfer.
When the transfer distance becomes large, we should enlarge evolution time $%
\tau $ to enhance the adiabaticity. In Fig.~\ref{fig4}(e), transfer fidelity as a
function of $J_{0}$ for transfer distance $D=21$ and evolution time $\tau
=1200/J$. The time evolution of the probabilities the same as Fig.~\ref{fig4}(b)-(d)
for $D=21$ and three different $J_{0}$ are illustrated in Fig.~\ref{fig4}(f)-(g). We
can still see from Fig.~\ref{fig2} that the optimum value of $J_{0}$ to achieve
high-fidelity transfer decreases as the transfer distance increase which is
consistent with the results shown before.

\begin{figure}[tbp]
\includegraphics[ bb=46 372 377 794, width=7 cm, clip]{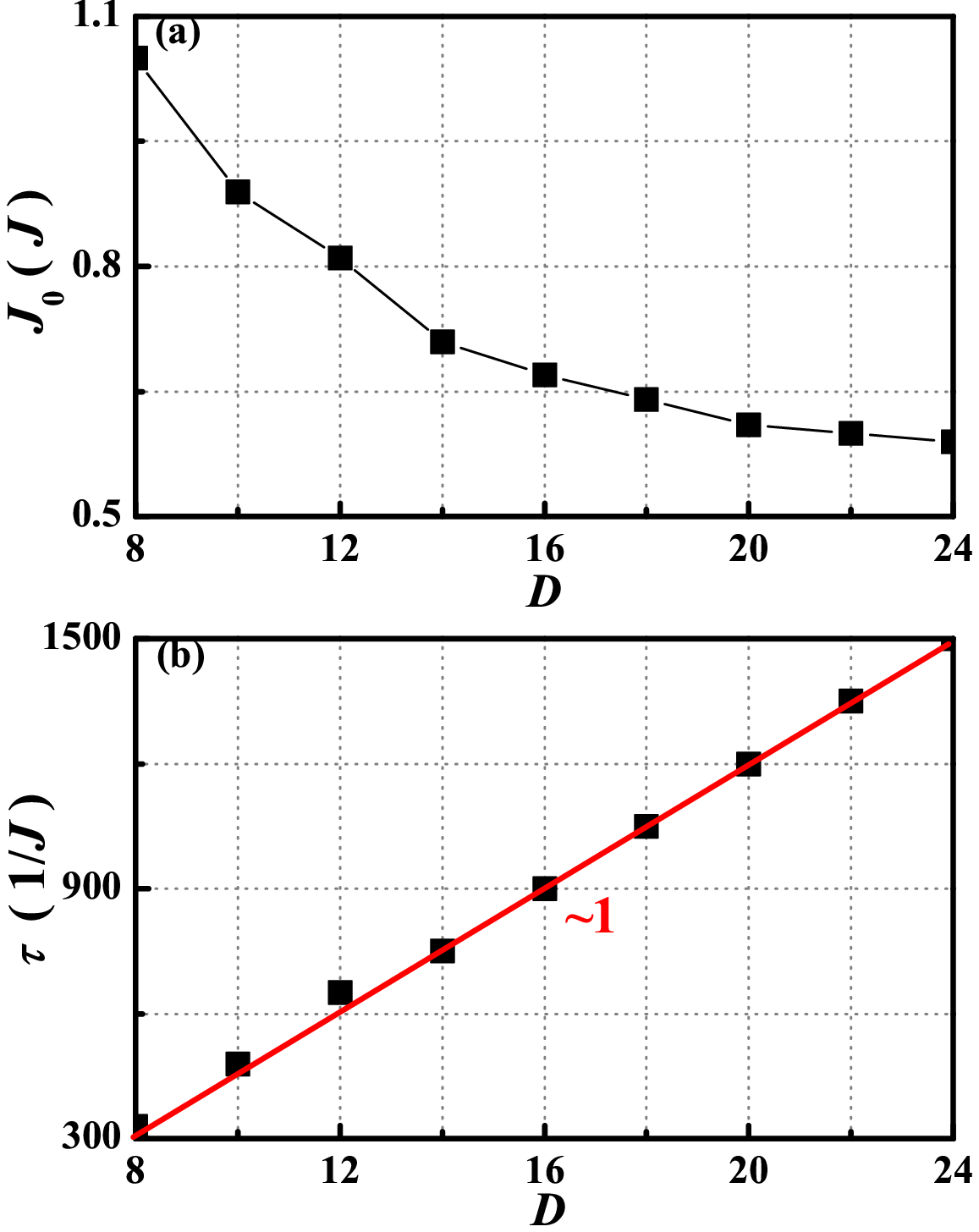}
\caption{(Color online) For a given chain length $N=48$ and a chosen fidelity $F=0.995$,
(a) the optimum value of coupling strength $J_{0}$ (in units of $J$) and (b)
the minimum time $\protect\tau$ (in units of $1/J$) required for state
transfer varies as a function of transfer distance $D$. The other system
parameters we choose is $J=1$, $\protect\mu_{0}=20$ and $\protect\alpha=8/%
\protect\tau$. It shows that optimum value of $J_{0}$ decreases as $D$
increasing and time $\protect\tau$ scales linearly with transfer distance $D$%
. }
\label{fig5}
\end{figure}

In order to provide the most economical choice of parameters for reaching
high transfer efficiency, we perform numerical analysis, as shown in Fig.~\ref{fig5}.
Specifically, we depict the optimum coupling $J_{0}$ (shown in Fig.~\ref{fig5}(a))
and the corresponding minimum state transfer time $\tau $ (shown in Fig.~\ref{fig5}(b))
for a given chain length $N=48$ and a given tolerable transfer
fidelity $F=99.5\%$, the minimum time varies as a function of transfer
distance $D$. One can see that the time $\tau $ required for high-fidelity
transfer scales linearly with transfer distance $D$.

\section{Summary and Discussion}

An efficient QST scheme should not only admit a state transfer of any
quantum state in a fixed period of time of the state evolution with high
fidelity, but also the transfer time can not grow fast as communication
distance increases. In this paper, we have introduced a long-range transport
mechanism for quantum information around a quasi-one-dimensional QDs
network, based on adiabatic passage. This scheme is realized by modulation
of gate voltages applied on the two external QDs which is connected to the
tight-binding chain. Under suitable system parameters, the electron can be
transported from the sender QD to the receiver one with high efficiency,
carrying along with it the quantum information encoded in its spin.
Different from the CTAPn Scheme~\cite{CTAPn}, our method is to induce
population transfer through tight-binding chain by maintaining the system in
its ground state and this is more operable in experiments. We have studied
the adiabatic QST through the system by theoretical analysis and numerical
simulations of the ground state evolution of tight-binding model. The result
demonstrates that it is an efficient high-fidelity process (99.5\%) for an
economical choice of system parameters. Increasing the transfer distance, we
found that the efficiency of QST is inversely proportional to the distance
of the two QDs.

In a real system, decoherence is the main obstacle to the experimental
implementation of quantum information~\cite{Ivanov,Kam}. There are two
sources of quantum decoherence in QDs, one
is due to charge dephasing brought by lead-QD coupling and the other is due
to the hyperfine interaction.
For the former, the coherence time of quantum dot is $%
\sim $1 ns, which plays a role in this case. On the other hand, the maximum
time in $N=48$ QD system needed for the appearance of the better fidelity is
roughly proportional to $10^{4}/J$. As a simple estimate of the effects of
decoherence, we compare this time with the dephasing time, which leads to a
limit of coupling strength of $J$ of $\sim $10 THz. The probability of
realization of this idea in experiment can be maximized by more precise
manipulation technology and by cooling the system. Furthermore, the development
of cold atom physics provides us with an alternative realization of our systems in experiment, 
because decoherence in cold-atom system is much less destructive.

\section*{Acknowledgments}

We acknowledge the support of the NSF of China (Grant No.10847150 and
No.11105086), the National Research Foundation and Ministry of Education,
Singapore (Grant No. WBS: R-710-000-008-271), the Shandong Provincial
Natural Science Foundation (Grant No. ZR2009AM026 and BS2011DX029), and the
basic scientific research project of Qingdao (Grant No.11-2-4-4-(6)-jch). Y.
X. also thanks the Basic Scientific Research Business Expenses of the
Central University and Open Project of Key Laboratory for Magnetism and
Magnetic Materials of the Ministry of Education, Lanzhou University (Grant
No. LZUMMM2011001) for financial support.


\begin{thebibliography}{99}
\bibitem{Bose1} S. Bose, Phys. Rev. lett. \textbf{91}, 207901 (2003).

\bibitem{Song} Z. Song and C. P. Sun, Low Temperature Physics \textbf{31},
686 (2005).

\bibitem{Christandle1} M. Christandl, N. Datta, A. Ekert and A.J. Landahl,
Phys. Rev. Lett. \textbf{92}, 187902 (2004).

\bibitem{Eckert} K. Eckert, M. Lewenstein, R. Corbal\'{a}n, G. Birkl, W.
Ertmer, and J. Mompart, Phys. Rev. A 70, 023606 (2004).

\bibitem{CTAP} A. D. Greentree, J. H. Cole, A. R. Hamilton, and L. C. L.
Hollenberg, Phys. Rev. B \textbf{70}, 235317 (2004).

\bibitem{STIRAP} N. V. Vitanov, T. Halfmann, B. W. Shore, and K. Bergmann,
Annu. Rev. Phys. Chem. \textbf{52}, 763 (2001).

\bibitem{atom1} K. Eckert, J. Mompart, R. Corbalan, M. Lewenstein, and G.
Birkl, Opt. Commun. \textbf{264}, 264 (2006).

\bibitem{atom2} T. Opatrn\'{y}, K. K. Das, Phys. Rev. A \textbf{79}, 012113
(2009).

\bibitem{spin} T. Ohshima, A. Ekert, D. K. L. Oi, D. Kaslizowski, L. C.
Kwek, e-print arXiv:quant-ph/0702019.

\bibitem{electron1} P. Zhang, Q. K. Xue, X. G. Zhao, and X. C. Xie, Phys.
Rev. A \textbf{69}, 042307 (2004).

\bibitem{electron2} J. Fabian and U. Hohenester, Phys. Rev. B \textbf{72},
201304(R) (2005).

\bibitem{BEC1} E. M. Graefe, H. J. Korsch, and D. Witthaut, Phys. Rev. A
\textbf{73}, 013617 (2006).

\bibitem{BEC2} M. Rab, J. H. Cole, N. G. Parker, A. D. Greentree, L. C. L.
Hollenberg, and A. M. Martin, Phys. Rev. A \textbf{77}, 061602(R) (2008).

\bibitem{BEC3} V. O. Nesterenko, A. N. Nikonov, F. F. de Souza Cruz, and E.
L. Lapolli, Laser Phys. \textbf{19}, 616 (2009).

\bibitem{LR1} L. C. L. Hollenberg, A. D. Greentree, A. G. Fowler, and C. J.
Wellard, Phys. Rev. B \textbf{74}, 045311 (2006).

\bibitem{LR2} A. D. Greentree, S. J. Devitt, and L. C. L. Hollenberg, Phys.
Rev. A \textbf{73}, 032319 (2006).

\bibitem{chen1} B. Chen, W. Fan, and Y. Xu, Phys. Rev. A \textbf{83}, 014301
(2011).

\bibitem{chen2} B. Chen, W. Fan, and Y. Xu, Sci China Ser G-Phys Mech
Astron, in press.

\bibitem{CF1} J. Cheng and J.-Y. Zhou, Phys. Rev. A \textbf{64}, 065402
(2001).

\bibitem{CF2} D. Goswami, Phys. Rep. \textbf{374}, 385 (2003).

\bibitem{Sun} D. Z. Xu, H. Lan, T. Shi, H. Dong, and C. P. Sun, Sci China
Ser G-Phys Mech Astron, \textbf{53}(7): 1234-1238 (2010).

\bibitem{CTAPn} L. C. L. Hollenberg, A. D. Greentree, A. G. Fowler, and C.
J. Wellard, Phys. Rev. B \textbf{74}, 045311 (2006).

\bibitem{Ivanov} P. A. Ivanov, N. V. Vitanov, and K. Bergmann, Phys. Rev. A
\textbf{70}, 063409 (2004).

\bibitem{Kam} I. Kamleitner, J. Cresser, and J. Twamley, Phys. Rev. A
\textbf{77}, 032331 (2008).

\bibitem{SZ} Z. Song, P. Zhang, T. Shi and C.-P. Sun, Phys. Rev. B \textbf{71%
}, 205314 (2005).
\end{thebibliography}
\end{document}